\documentclass[preprint,review,12pt]{elsarticle}
\usepackage{amssymb}
\usepackage{array}
\usepackage{marginnote}


\usepackage{lineno}
\usepackage{graphicx}
\usepackage{subfigure}
\usepackage{multirow}
\usepackage{amsmath}
\usepackage{graphicx}
\usepackage[labelsep=period]{caption}
\usepackage[labelfont=bf]{caption}

\usepackage[]{hyperref}
\usepackage[utf8]{inputenc}
\UseRawInputEncoding
\hypersetup{hidelinks}
\journal{NIMA}
\begin{document}
\begin{frontmatter}

\title{\boldmath Design and testing of LGAD sensor with shallow carbon implantation}
\author[label1,label2,label3]{Kewei Wu\fnref{kewei}}
\author[label1,label2,label3]{Xuewei Jia\fnref{xuewei}}
\author[label1,label2,label3]{Tao Yang}
\author[label1,label2,label3]{Mengzhao Li}
\author[label1,label3]{Wei Wang}

\author[label1,label3]{Mei Zhao\corref{cor}}
\ead{zhaomei@ihep.ac.cn}
\author[label1,label3]{Zhijun Liang\corref{cor}}
\ead{liangzj@ihep.ac.cn}
\author[label1]{Jo\~{a}o Guimar\~{a}es da Costa}

\author[label1,label3]{Yunyun Fan}
\author[label1,label2,label3]{Han Cui}

\author[label5]{Alissa Howard}
\author[label5]{Gregor Kramberger}

\author[label1,label3]{Xin Shi}
\author[label1,label2,label3]{Yuekun Heng}

\author[label1,label2,label3]{Yuhang Tan}
\author[label1,label3]{Bo Liu}
\author[label1,label2,label3]{Yuan Feng}
\author[label1,label2,label3]{Shuqi Li}
\author[label1,label2,label3]{Mengran Li}
\author[label1,label2,label3]{Chengjun Yu}
\author[label1,label3]{Xuan Yang}
\author[label1,label2,label3]{Mingjie Zhai}

\author[label4]{Gaobo Xu}
\author[label2,label4]{Gangping Yan}
\author[label2,label4]{Qionghua Zhai}
\author[label4]{Mingzheng Ding}
\author[label4]{Jun Luo}
\author[label4]{Huaxiang Yin}
\author[label4]{Junfeng Li}

\address[label1]{Institute of High Energy Physics, Chinese Academy of Sciences, 19B Yuquan Road, Shijingshan, Beijing 100049, China}
\address[label2]{University of Chinese Academy of Sciences, 19A Yuquan Road, Shijingshan, Beijing 100049, China}
\address[label3]{State Key Laboratory of Particle Detection and Electronics, 19B Yuquan Road, Shijingshan, Beijing 100049, China}
\address[label5]{Jožef Stefan Institute, Jamova 39, SI-1000 Ljubljana, Slovenia}
\address[label4]{Institute of Microelectronics, Chinese Academy of Sciences, 3 Beitucheng West Road, Chaoyang, Beijing 100029, China}

\cortext[cor]{Corresponding Author}

\begin{abstract}
The low gain avalanche detectors (LGADs) are thin sensors with fast charge collection which in combination with internal gain deliver an outstanding time resolution of about 30 ps.
High collision rates and consequent large particle rates crossing the detectors at the upgraded Large Hadron Collider (LHC) in 2028 will lead to radiation damage and deteriorated performance of the LGADs.
The main consequence of radiation damage is loss of gain layer doping (acceptor removal) which requires an increase of bias voltage to compensate for the loss of charge collection efficiency and consequently time resolution.
The Institute of High Energy Physics (IHEP), Chinese Academy of Sciences (CAS) has developed a process based on the Institute of Microelectronics (IME), CAS capability to enrich the gain layer with carbon to reduce the acceptor removal effect by radiation.
After 1 MeV neutron equivalent fluence of 2.5$\times$10$^{15}$ n$_{eq}$/cm$^{2}$, which is the maximum fluence to which sensors will be exposed at ATLAS High Granularity Timing Detector (HGTD), the IHEP-IME second version (IHEP-IMEv2) 50 $\mu$m LGAD sensors already deliver adequate charge collection \textgreater4 fC and time resolution \textless50 ps at voltages \textless400 V. The operation voltages of these 50 $\mu$m devices are well below those at which single event burnout may occur.
\end{abstract}

\end{frontmatter}


\section{Introduction} 

In order to exploit its full potential, the Large Hadron Collider (LHC) will be upgraded to achieve larger luminosity (High Luminosity-LHC) in 2028. 
The instantaneous luminosity will reach levels exceeding the present ones by at least a factor of five~\citep{HL-LHC_BOOK}.

Low gain avalanche detector (LGAD) sensors are thin ($\approx$~50~$\mu$m) silicon sensors (structure: n$^{++}$/p$^{+}$/p$^{-}$/p$^{++}$) with outstanding time resolutions ($\approx$~30~ps) and moderate gains (\textless ~100). 
With robust performance in irradiation environments, it is feasible to use LGAD sensors in the HL-LHC. 
The Centro Nacional de Microelectr\'{o}nica (CNM) Barcelona made the first developments and measurements with LGAD sensors~\citep{pellegrini2014LGAD}, which have been followed by many others~\citep{SADROZINSKI_2014_UFSD, giacomini2019BNL, wada2019HPK, carulla2019CNM50mum, ferrero2019FBKradiation_resistant}.
The key property of an LGAD sensor is a gain layer (at n$^{++}$/p$^{+}$ junction) that is carefully tuned to give sufficient gain at moderate voltages at which the active thickness of the LGAD can be depleted to achieve high drift velocities. 
Radiation mainly affects the active doping concentration of boron atoms in the gain layer through the so-called acceptor removal mechanism \citep{Moll_Thesis} which in turn deteriorates the time resolution of LGAD sensors.

The rate at which this process occurs depends on several factors, mainly the initial concentration of boron atoms and added impurities to the gain layer.
An active acceptor deactivates exponentially as a function of fluence (e.g., Equation~\eqref{eq:a}):
\begin{equation}
\label{eq:a}
N_{boron}\left(\phi_{eq}\right)=N_{boron}(0) e^{-c\cdot \phi_{eq}},
\end{equation}
where $c$ is the acceptor removal constant, $N_{boron}$ is the amount of active boron atoms in the gain layer, and $\phi_{eq}$ is the equivalent fluence of 1~MeV neutrons.

The usage of carbon implantation to slow down the acceptor removal effect in LGAD sensors was first proposed by \cite{gregor_carbon_idea} and realized by the Fondazione Bruno Kessler (FBK) \citep{ferrero2019FBKradiation_resistant}. 
The implanted carbon competes with acceptors to form ion--carbon complexes instead of ion--acceptor complexes.
In previous years, several studies indicated carbon could be the main component to reduce the removal constant $c$ \citep[e.g.,]{ARCIDIACONO2020FBK}. 
This strategy was also adopted in the Institute of High Energy Physics (IHEP) and the Institute of Microelectronics (IME) second version (IHEP-IMEv2) sensor design.
LGAD sensors with carbon implantation have disadvantages such as boron deactivation by carbon and increased leakage currents \citep{firstFBKproduction}. 
The IHEP-IMEv2 LGAD sensor, with low energy carbon implantation, has a stable gain layer in which the active boron concentration does not decrease with increasing carbon dose.
Meanwhile, the IHEP-IMEv2 carbon-enriched LGAD sensor has the smallest acceptor removal constant among the IHEP-IME first version (IHEP-IMEv1) \citep{Mengzhaov1}, FBK Ultra-Fast Silicon Detectors third version (UFSD3) \citep{ARCIDIACONO2020FBK}, and Hamamatsu Photonics Kabushiki-gaisha 3.2 version (HPK3.2) \citep{HPK_Acceptor_Removal_UFSD} LGAD sensors, as shown in Table~\ref{tab:0}. 
This is reflected in the smallest excess voltage required to compensate for the gain loss due to radiation.

\begin{center}
\begin{table}
\captionsetup{justification=raggedright,labelsep=newline,singlelinecheck=off,margin=20pt,}
	\caption{\label{tab:0} Acceptor removal constants of different LGAD sensors.}
	\centering
\begin{tabular}{m{1.5cm}<{\centering}m{6cm}<{\centering}m{4cm}<{\centering}}
	\hline
	\centering
	Producer&Type&Acceptor removal constant / 10$^{-16}$ cm$^{2}$ \\
	\hline
	IHEP&IHEP-IMEv2 Wafer7 second quadrant (W7-II)&1.27\\
	IHEP&IHEP-IMEv1 Wafer1 forth quadrant (W1-IV)&3.12\\
	\hline
	FBK& UFSD3, Epitaxial substrate Wafer 4, Low Diffusion Boron gain layer, $\mathrm{C}_{A}$ dose of carbon (UFSD3\underline{  }B LD$+$C\underline{  }A Epi(W4)) &1.45\\
	FBK& Wafer 7, Low Diffusion Boron gain layer, $\mathrm{C}_{B}$ dose of carbon (UFSD3\underline{  }B LD$+$C\underline{  }B(W7)) &2.48\\
	FBK&UFSD2, Wafer 1, Low Diffusion Boron gain layer, no carbon (UFSD2\underline{  }B LD(W1)) &4.66\\
	\hline
    HPK&HPK3.2&3.15\\
    HPK&HPK3.1&5.20\\
	\hline
\end{tabular}
\end{table}
\end{center}

The total radiation goals of the ATLAS High Granularity Timing Detector (HGTD) \citep{ATLAS_HGTD_TDR} and the CMS Minimum Ionizing Particles (MIP) Timing Detector (MTD) Endcap Timing Layer (ETL) \citep{CMS_MIP_TDR} are 2.5$\times$10$^{15}$ n$_{eq}$/cm$^{2}$ and 1.5$\times$10$^{15}$ n$_{eq}$/cm$^{2}$ (i.e., a 1 MeV neutron-equivalent fluence), respectively, implying \textless 2 MGy total ionizing dose (TID) for both.

This paper shows the efforts of the IHEP to solve the abovementioned irradiation issues by implanting carbon into LGAD with low energy (shallow carbon). 
A careful investigation of processing parameters resulted in LGAD sensors that fulfil the radiation hardness requirements for application at the HL-LHC. 
The sensor's characteristics before and after irradiation are presented.

\section{LGAD design fabrication}

IHEP has developed two versions of LGAD sensors, based on the Institute of Microelectronics (IME) process capability, which were irradiated at the Jozef Stefan Institute (JSI) research reactor with neutrons to evaluate the sensors' performance after irradiation.

The first version of LGAD sensors that IHEP developed with IME focused on optimizing the n$^{++}$ layer and the p$^{+}$ gain layer fabrication process \citep{Kewei_v1,YangTao_v1}. 
A preliminary exploration of using shallow carbon implantation in the LGAD sensors was also included in IHEP-IMEv1 \citep{Mengzhaov1,ZhaoMei_v1}.
LGAD sensors with carbon implantation showed better performance than non-carbon-implanted LGAD sensors with the same n$^{++}$ layer and the p$^{+}$ gain layer after irradiation of 2.5$\times$10$^{15}$ n$_{eq}$/cm$^{2}$. 
The second version of LGAD sensors developed by IHEP with IME (IHEP-IMEv2) focused on optimizing the carbon implantation and annealing process to improve their radiation hardness.

\begin{figure}[!t]
	\centering
	\includegraphics[width=0.59\textwidth]{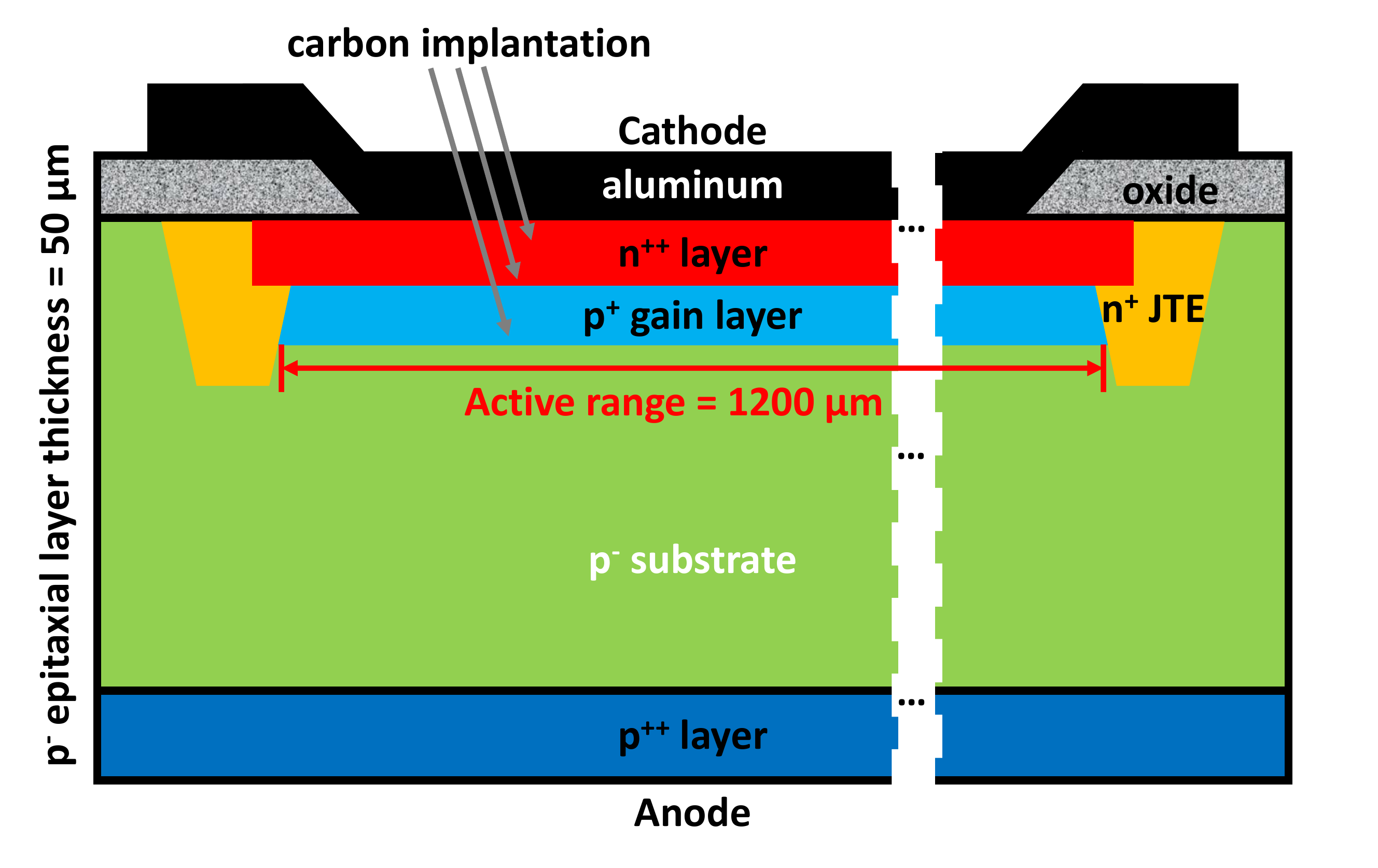}
	\caption{\label{fig:11} Sketch of the LGAD structure with the active area shown. The height and width are not to scale. The sensor total area is 1200 $\mu$m $\times$ 1200 $\mu$m. The thickness of the p$^{-}$ epitaxial layer (p$^{-}$ substrate) is 50 $\mu$m. The carbon atoms were implanted in the n$^{++}$ layer which diffused to the p$^{+}$ gain layer.}
\end{figure}

\begin{figure}[htbp]
	\centering
	\includegraphics[width=0.49\textwidth]{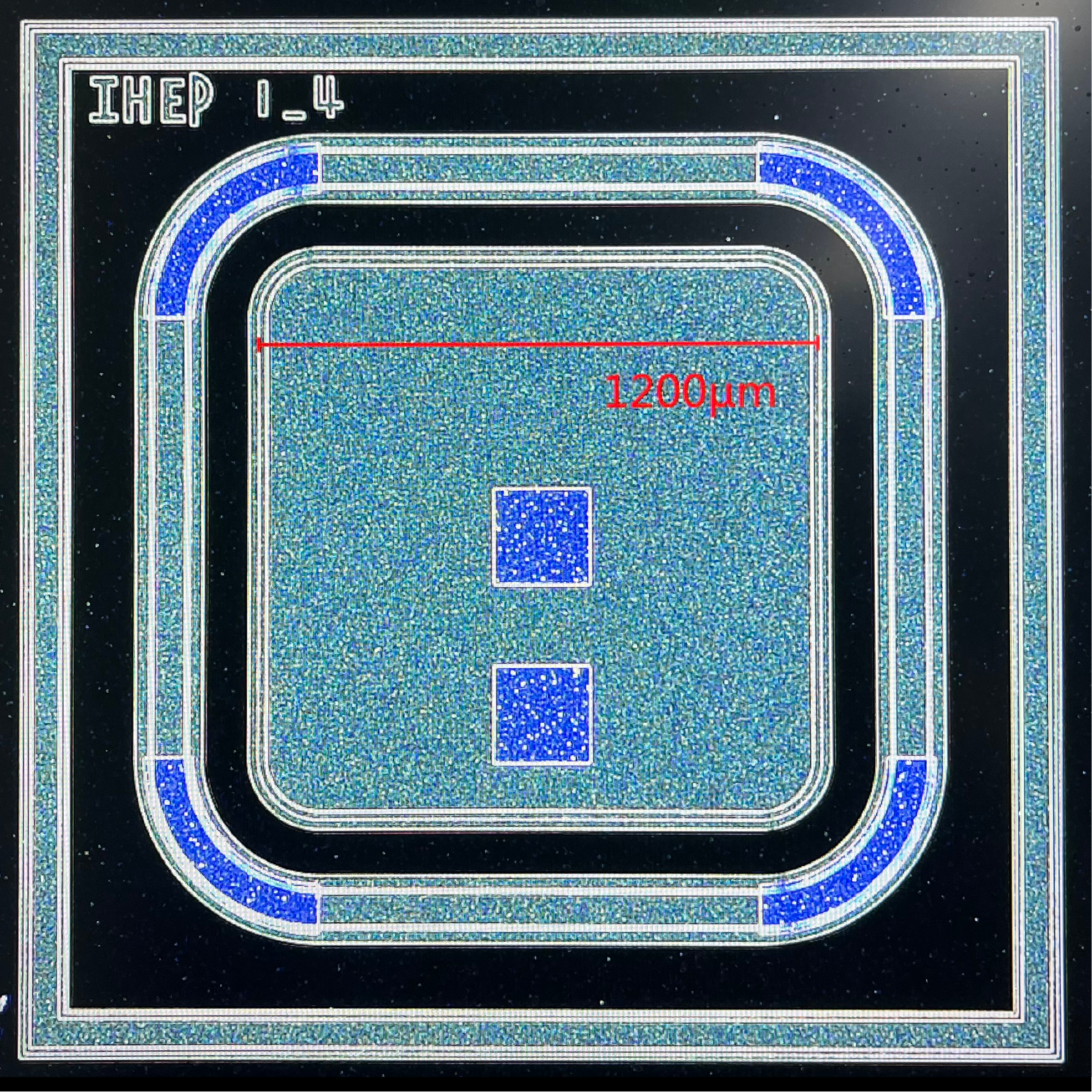}
	\caption{\label{fig:12} Picture of a single-pad IHEP-IMEv2 LGAD sensor under a microscope. The active range of 1200 $\mu$m has been labeled by the red line. The metal contact pads were shown in blue.}
\end{figure}

Both the IHEP-IMEv1 and IHEP-IMEv2 LGAD sensors were fabricated with an n$^{++}$/p$^{+}$/p$^{-}$/p$^{++}$ structure on silicon wafers with a 50 $\mu$m thick high-resistivity epitaxial layer, as shown in Fig.~\ref{fig:11}. 
These devices are reversed biased, with a high voltage applied to the p$^{++}$ layer anode.  
Electrons drift to the p$^{+}$ gain layer, where a high electric field multiplies them, and then drift to the n$^{++}$ cathode at ground potential. A photo of a single pad device is shown in Fig.~\ref{fig:12}.

\begin{figure}[htbp]
	\centering
	\includegraphics[width=0.49\textwidth]{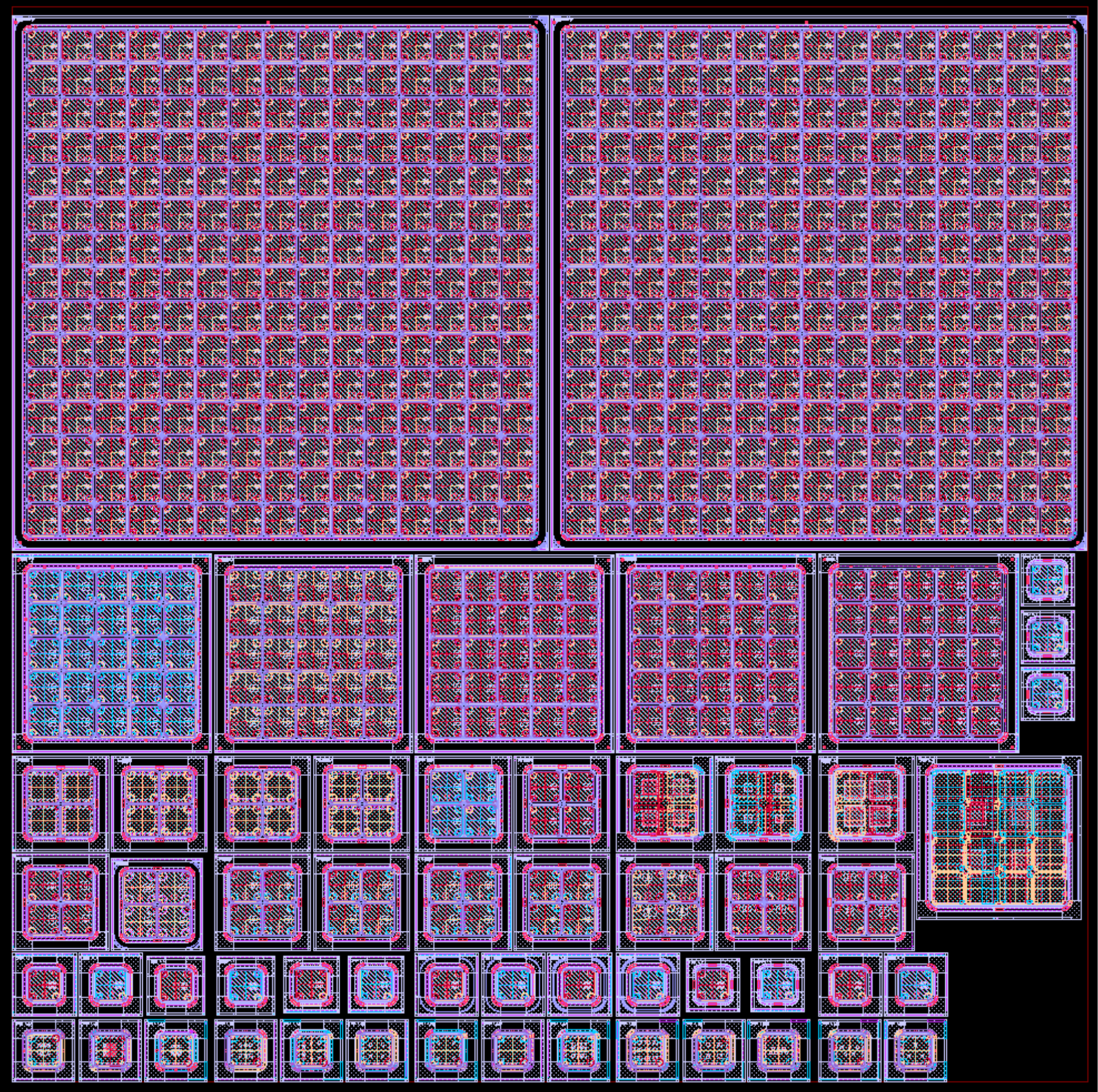}
	\caption{\label{fig:21} IHEP-IMEv2 $4\times4$ cm$^{2}$ layout mask. Sizes of $15\times15$, $5\times5$, $2\times2$, and single pads were allocated and aligned in this mask.}
\end{figure}

In both the IHEP-IMEv1 and IHEP-IMEv2 sensor production, the wafers were split into four quadrants. 
The devices for each reticle include single pads, $2\times2$, $5\times5$, and $15\times15$ (full size) LGAD and PIN sensors, as shown in Fig.~\ref{fig:21}. 
The PIN sensors are identical to the LGAD sensors but without a p$^{+}$ gain layer.

Apart from the carbon process, the IHEP-IMEv2 carbon-enriched LGAD sensors inherited the boron and phosphorus process from the IHEP-IMEv1 W7-IV recipe \citep{Kewei_v1,YangTao_v1}.
IHEP-IMEv2 has an independent carbon process that implants and diffuses carbon into the active area before the p$^{+}$ layer and n$^{++}$ layer processes. 
This minimizes the carbon influence on changing the boron and phosphorus doping profile. 
The radiation hardness of the LGAD sensors was optimized using the different carbon processes shown in Table~\ref{tab:a}, where the relative carbon doses are expressed in arbitrary units (a.u.). 
Three wafers underwent carbon implantation. 
Wafer 4 received four different carbon doses in four quadrants that went through a fast thermal process, while wafer 7 and wafer 8 received eight different carbon doses in eight quadrants that went through a long-time thermal process.

\begin{center}
\begin{table}
\captionsetup{justification=raggedright,
labelsep=newline,
singlelinecheck=off,margin=20pt,
}
	\caption{\label{tab:a} Key Parameters of the IHEP-IMEv2 LGAD sensors. The relative carbon doses are expressed in arbitrary units (a.u.).}
	\centering
\begin{tabular}{m{1cm}<{\centering}m{1.6cm}<{\centering}m{1.6cm}<{\centering}m{3.5cm}<{\centering}}
	\hline
	\centering
	Wafer&Quadrant&Carbon Dose&Carbon Thermal Process\\
	\hline
	4 & I & 0.2 a.u.& fast\\
	4 & II & 1 a.u.& fast\\
	4 & III & 5 a.u.& fast\\
	4 & IV & 10 a.u.& fast\\
	\hline
	7 & I & 0.2 a.u.& long-time\\
	7 & II & 0.5 a.u.& long-time\\
	7 & III & 1 a.u.& long-time\\
	7 & IV & 3 a.u.& long-time\\
	\hline
	8 & I & 6 a.u.& long-time\\
	8 & II & 8 a.u.& long-time\\
	8 & III & 10 a.u.& long-time\\
	8 & IV & 20 a.u.& long-time\\
	\hline
\end{tabular}
\end{table}
\end{center}

\section{Non-irradiated sensor characterization}
\subsection{Doping profile by SIMS}
\begin{figure}[!b]
	\centering
	\includegraphics[width=0.49\textwidth]{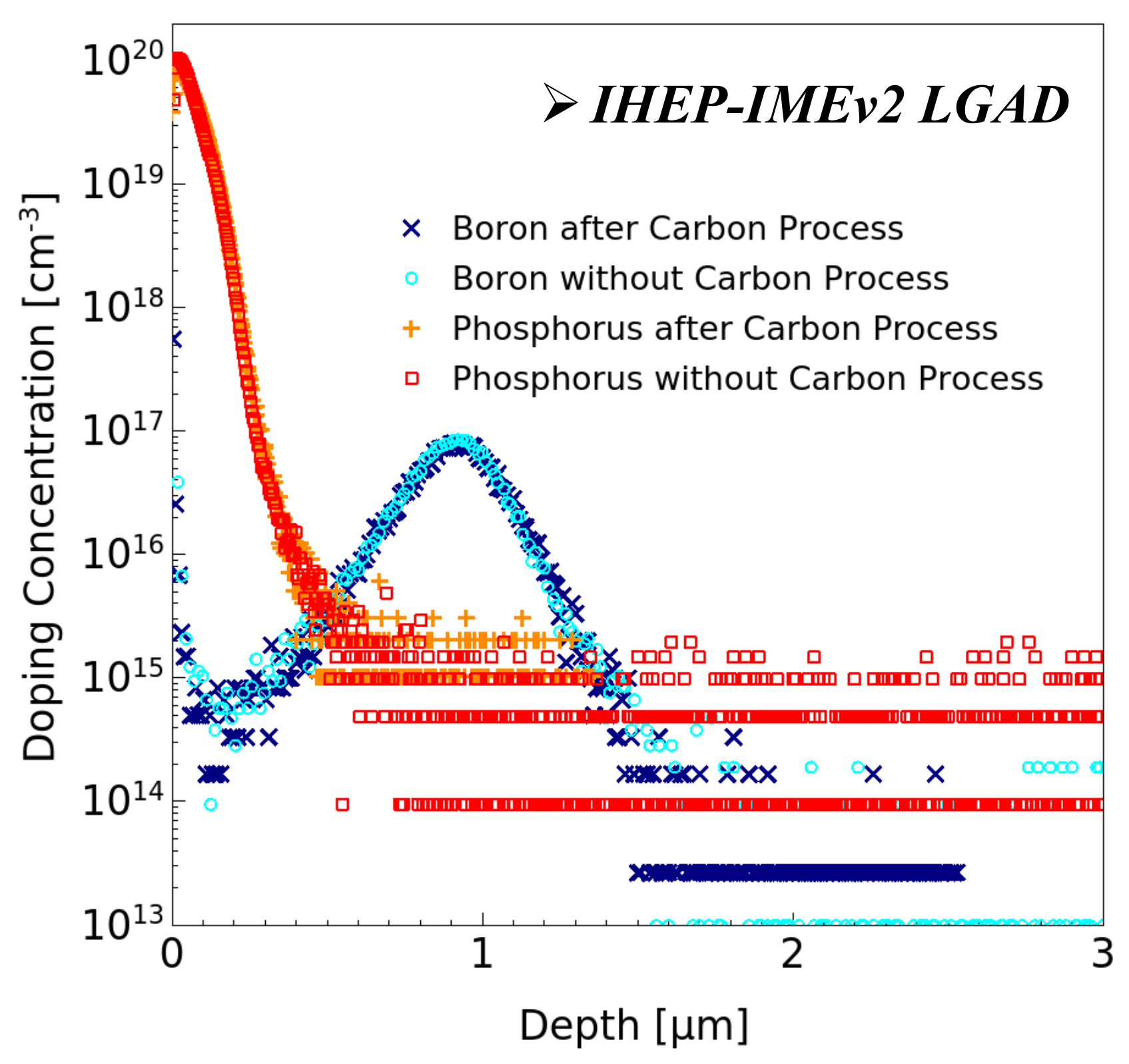}
	\caption{\label{fig:311} IHEP-IMEv1 and IHEP-IMEv2 doping profiles extracted from SIMS measurements. The IHEP-IMEv2 process, inherited from IHEP-IMEv1, shows good reproducibility. The minimum detection concentration of boron and phosphorus is 10$^{14}$~cm$^{-3}$ and 10$^{15}$~cm$^{-3}$, respectively.}
\end{figure}

The simulation-process parameters of IHEP \citep{YangTao_v1,ZhaoMei_v1} were calibrated according to the Secondary Ion Mass Spectrometry (SIMS) test results based on the stability of the IME process. 
The boron and phosphorus doping profiles in the IHEP-IMEv2 sensors, as measured by SIMS, are both highly consistent with those of the IHEP-IMEv1 non-carbon sensors (see Fig.~\ref{fig:311}).

\subsection{I--V and leakage current}
\begin{figure}[!b]
	\centering
	\includegraphics[width=0.99\textwidth]{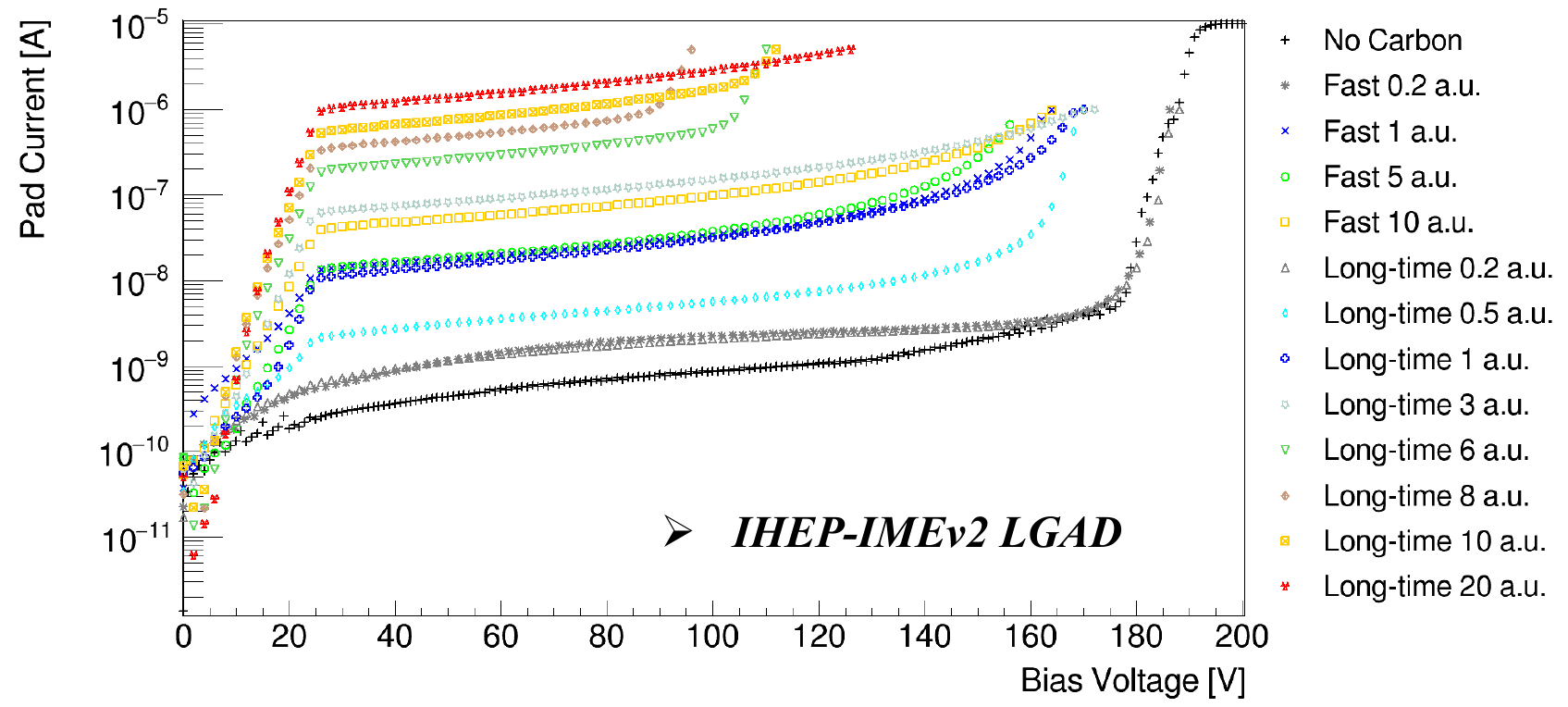}
	\caption{\label{fig:321} I--V test results of the LGAD sensors with different carbon doses (0.2--20 a.u.) and different thermal processes (fast and long-time). The breakdown voltages decrease with an increase of carbon dose, while the leakage current, at a given voltage, increases with carbon dose.}
\end{figure}

Current--voltage (I--V) tests were performed to find the breakdown voltage and leakage current levels. 
Different carbon doses (0.2--20 a.u.) and different thermal processes (fast and long-time) affect the I--V profile. 
Non-irradiated carbonated sensor I--V curves from IHEP-IMEv2 sensors are shown in Fig.~\ref{fig:321}.

\begin{figure}[!t]
	\centering
	\includegraphics[width=0.49\textwidth]{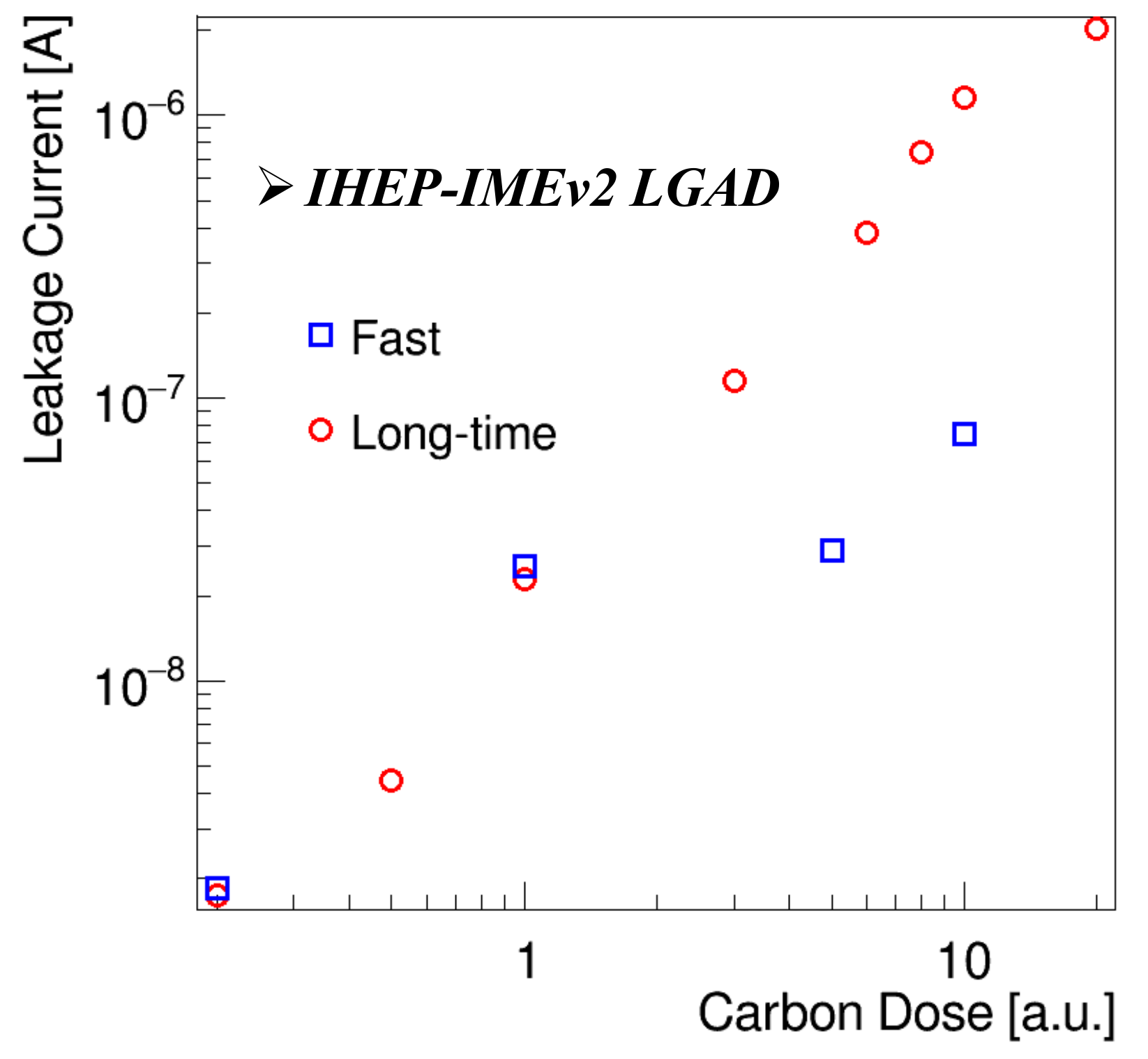}
	\caption{\label{fig:322} Leakage current (I$_{leak}$) of the LGAD sensors for different carbon doses (0.2--20~a.u.) and different thermal processes (fast and long-time) at 80~V. The long-time annealing process increases the leakage current level more than the fast annealing process at carbon doses larger than 1 a.u. The I$_{leak}$ value of the non-carbon sensor at 80~V is 0.68~nA.}
\end{figure}

The leakage current measured, at room temperature with bias voltage of -80 V ($\gg$ $V_{fd}$ of the device), versus carbon dose, is shown in Fig.~\ref{fig:322}. 
The leakage current significantly grows from 10$^{-9}$ A to 10$^{-6}$ A with increasing carbon dose. 
Clearly pointing to the formation of carbon-related defects giving rise to higher energy levels in the band-gap and consequently an increase in generation current.

\subsection{Capacitance--voltage and gain layer depletion voltage}
\begin{figure}[!t]
	\centering
	\includegraphics[height=0.37\textwidth]{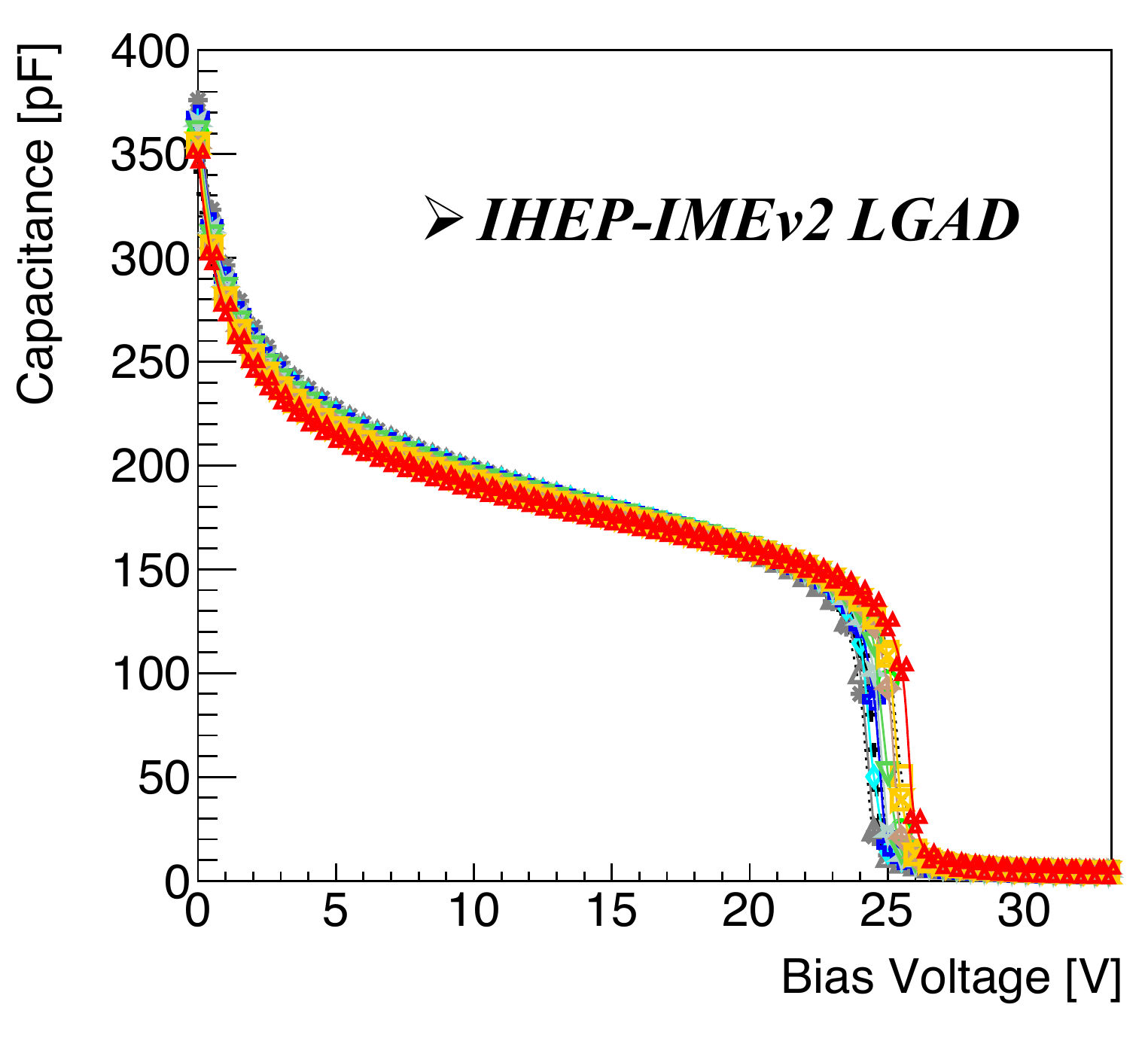}
	\includegraphics[height=0.37\textwidth]{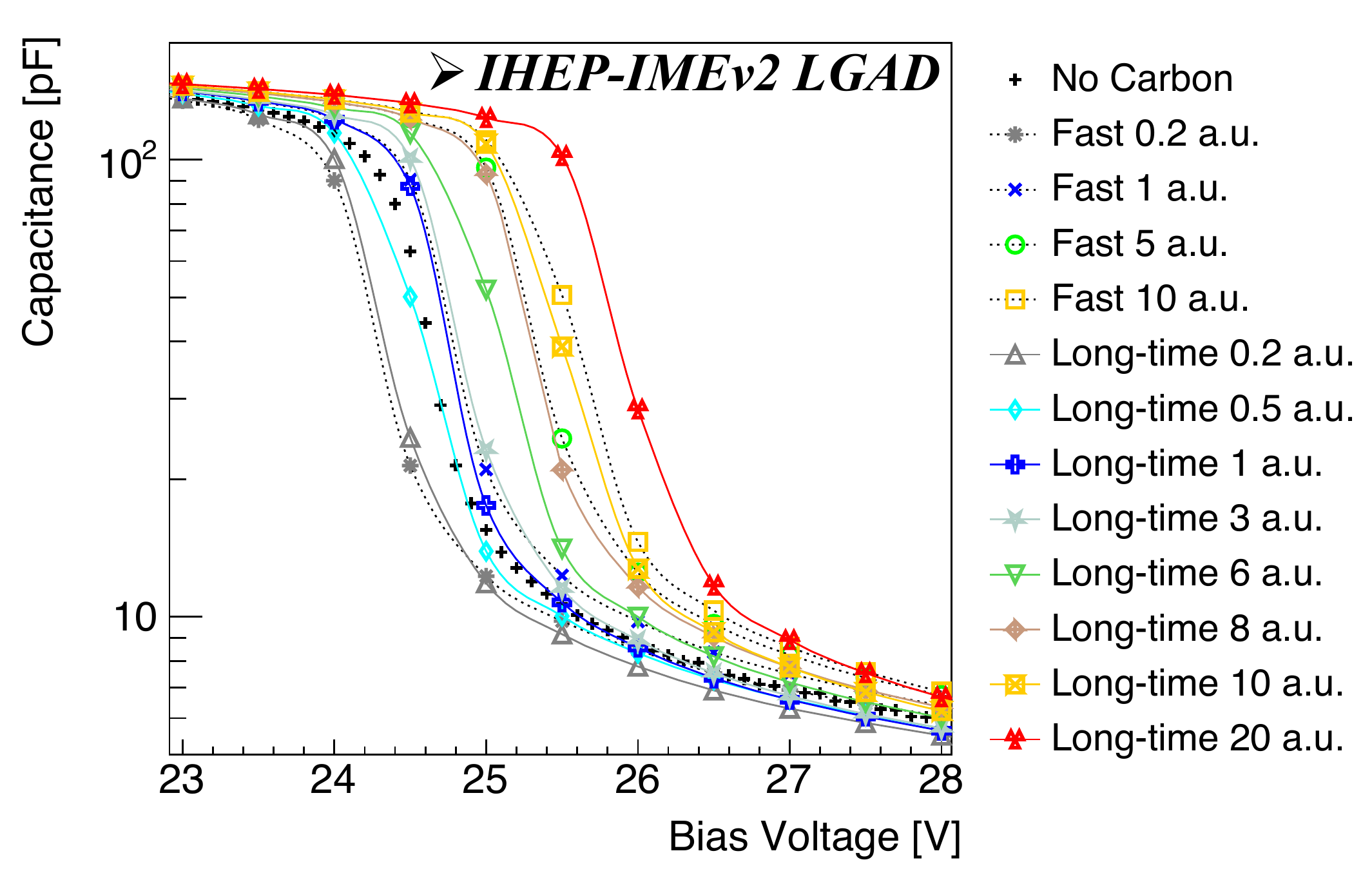}
	\caption{\label{fig:323} Left: C--V measurements of the LGAD sensors for different carbon doses (0.2--20 a.u.) and different thermal processes (fast and long-time) in the bias voltage range 0 to -30 V. Right: Amplified view of the C--V measurements around $V_{gl}$. It can be seen that $V_{gl}$ increases as the amount of carbon implantation increases.}
\end{figure}

Capacitance--voltage (C--V) tests were performed to find the gain layer depletion voltage ($V_{gl}$) and full depletion voltage ($V_{fd}$).
As seen in the I--V tests, different carbon doses (0.2--20 a.u.) and different thermal processes (fast and long-time) affect the C--V profile. Non-irradiated carbonated sensor C--V curves are shown in Fig.~\ref{fig:323}.

\begin{figure}[!b]
	\centering
	\includegraphics[width=0.49\textwidth]{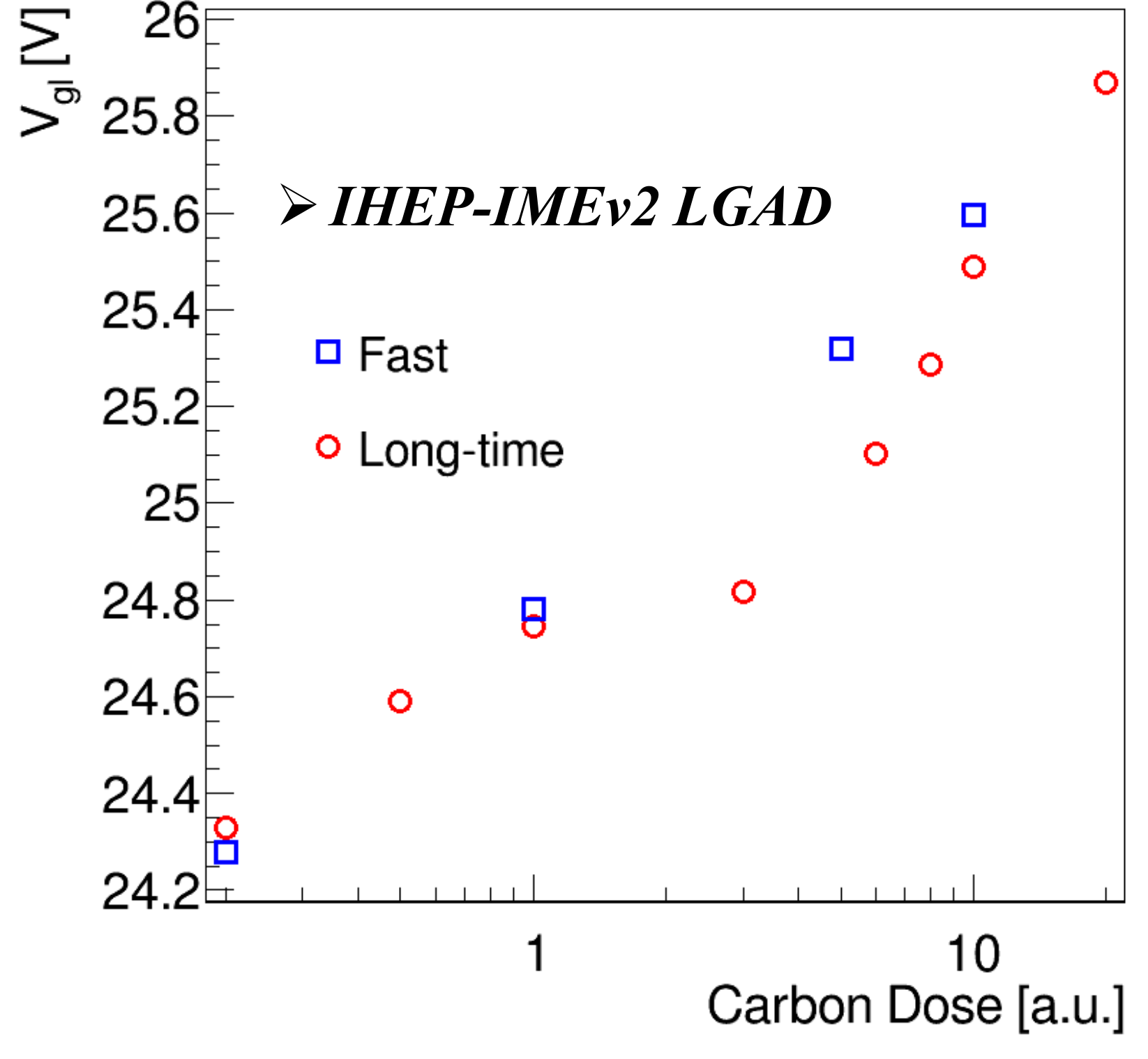}
	\caption{\label{fig:324} $V_{gl}$ values of the LGAD sensors, for both fast and long-time thermal processes, which increase as the carbon dose increases from 0.2 to 20 a.u. The range of $V_{gl}$ values spans about $\Delta$$V_{gl}$ $\approx$1.5~V.}
\end{figure}

The $V_{gl}$ voltages for all investigated samples are shown in Fig.~\ref{fig:324}. 
Unlike observations made for carbonated FBK sensors by \citep{firstFBKproduction}, here $V_{gl}$ is found to increase with carbon concentration.

The exact phenomena responsible for this behavior required further investigation. 
One relevant point here is that the IHEP-IMEv2 sensors have shallow carbon implantation, while the FBK sensors have a deeper implantation. 
Hence, a possible explanation is that, due to carbon deactivation of the effective dopant, the IHEP-IMEv2 sensors have fewer active donors in n$^{++}$ layer, while for the FBK sensors, there are fewer donors in the  p$^{+}$ gain layer.
In the case of IHEP-IMEv2, the reduction of active donors in the n$^{++}$ layer shifts the starting point of the space charge area to a shallower region, thereby increasing the total depletion depth, resulting in an increase in $V_{gl}$.

\section{Sensor performance after irradiation}
\subsection{Neutron irradiation and beta test setup}

\begin{figure}[!b]
	\centering
	\includegraphics[width=0.59\textwidth]{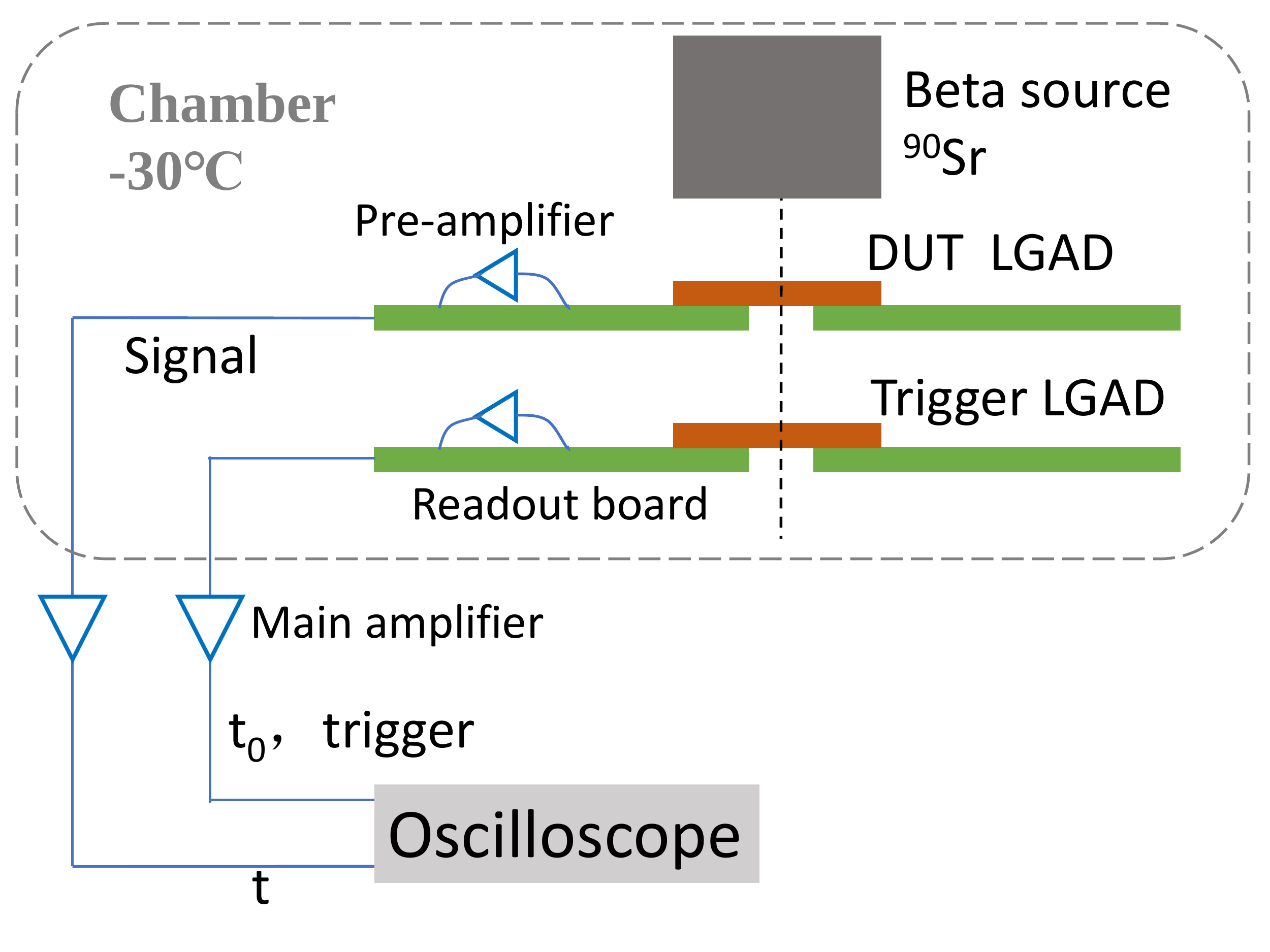}
	\caption{\label{fig:BetaSetup} Schematic diagram of IHEP-IME LGAD sensors beta telescope setup \citep{Mengzhaov1}.}
\end{figure}

The IHEP-IMEv2 and IHEP-IMEv1 LGAD sensors \citep{Mengzhaov1} were irradiated with neutrons at the JSI nuclear research reactor in Ljubljana \citep{JSI_Source,Padilla_2020_DeepGainLayer}. 
Three different irradiation fluences were used: 0.8$\times$10$^{15}$, 1.5$\times$10$^{15}$, and 2.5$\times$10$^{15}$ n$_{eq}$/cm$^{2}$.
The irradiated LGAD sensors were annealed at 60~$^{\circ}$C for 80~min before testing their performance with electrons from a Sr-90 source (beta source test). 
The beta source tests of the IHEP-IMEv2 LGAD sensors were performed in a climate chamber at -30~$^{\circ}$C with a Sr-90 radiation source. 
The LGAD sensors were wire bonded to readout boards, designed by the University of California Santa Cruz (UCSC), using wide bandwidth transimpedance SiGe amplifiers \citep{beamtest_16ps}.
Fig.~\ref{fig:BetaSetup} shows the beta telescope experimental setup.
The lower sensor is used as the trigger for electrons signal in the beta telescope test. 
The signal pulses from both sensors are recorded by a digital oscilloscope with 2~GHz bandwidth and 40~GS/s sampling rate for offline analysis \citep{Mengzhaov1}.

\begin{figure}[!b]
	\centering
	\includegraphics[scale=0.36]{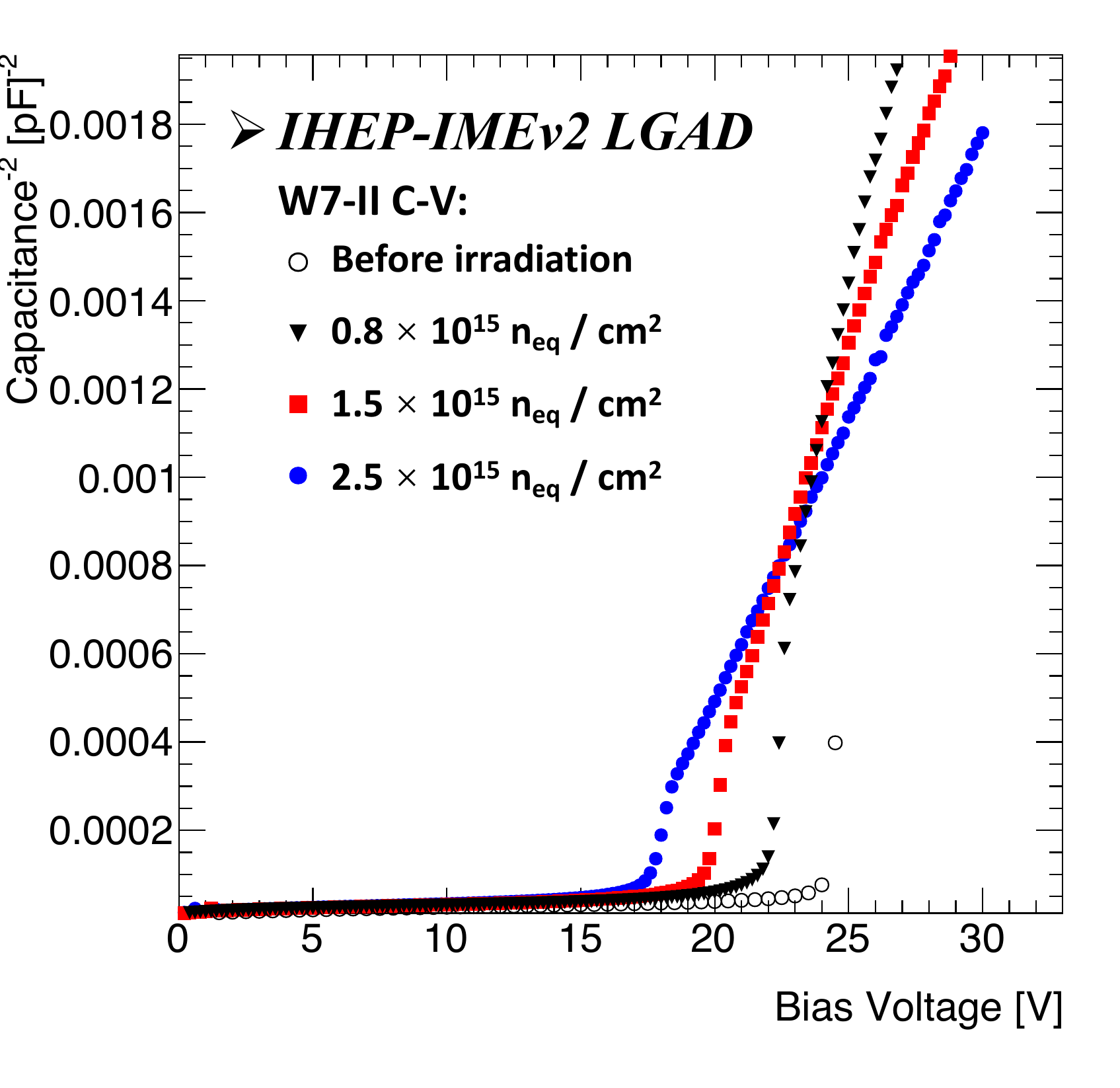}
	\includegraphics[scale=0.36]{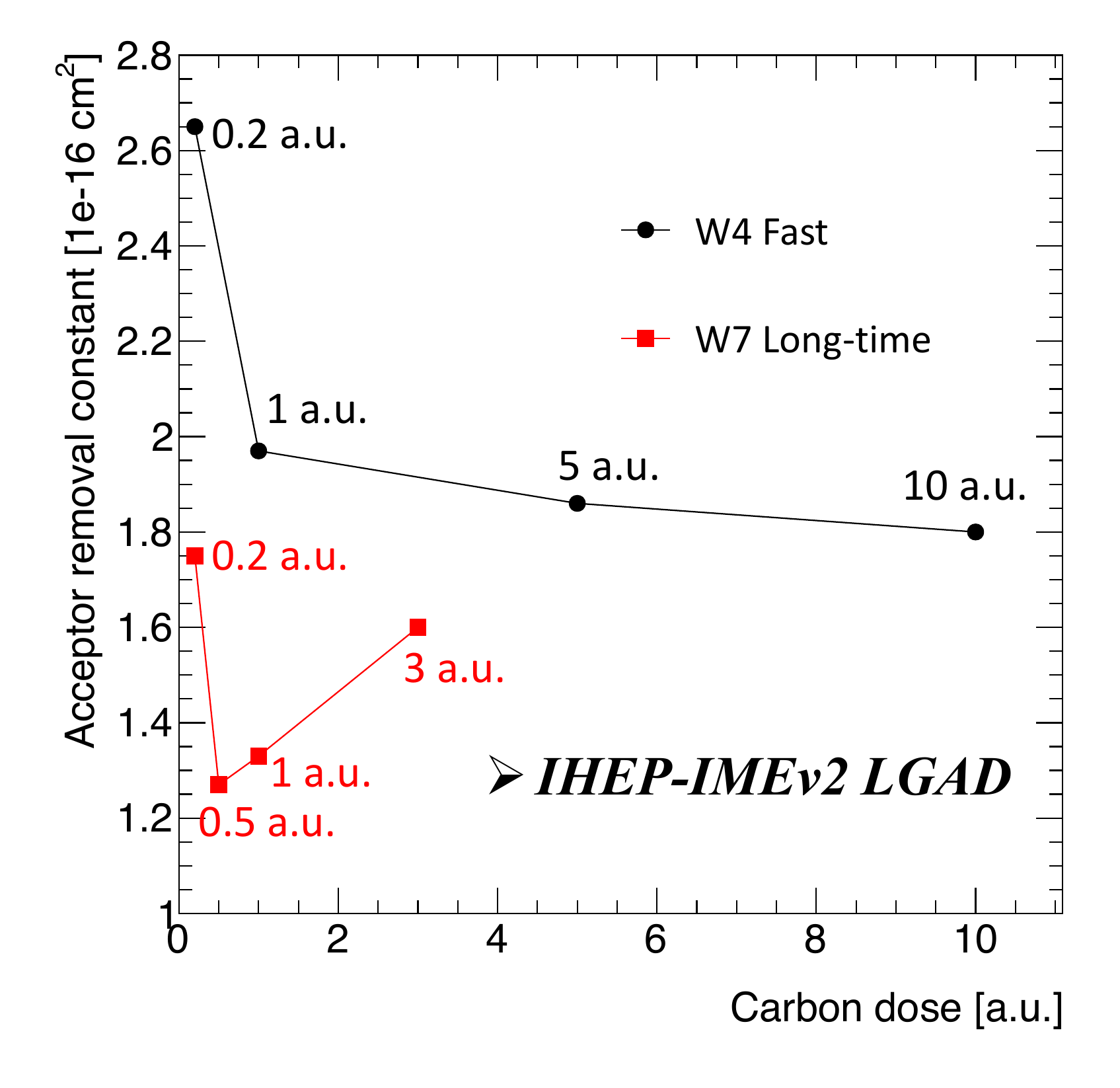}
	\caption{\label{fig:423} Left: C--V measurements of the IHEP-IMEv2 LGAD sensor with a 0.5~a.u. carbon dose and long-time carbon annealing recipe (W7-II) before and after receiving 0.8$\times$10$^{15}$, 1.5$\times$10$^{15}$, and 2.5$\times$10$^{15}$ n$_{eq}$/cm$^{2}$ radiation fluences. $V_{gl}$ decreases with increasing radiation fluence. \\Right: acceptor removal constants of IHEP-IMEv2 LGAD sensors for different carbon doses and two different thermal processes.} 
\end{figure}

\subsection{Acceptor removal constants} 
The acceptor removal constants (Equation~\eqref{eq:c}), which are reflected in the gain--loss, were extracted from the C--V measurements and calculated using the $V_{gl}$ voltages for the different radiation fluences.
Fig.~\ref{fig:423}(left) shows the C--V changes after the different irradiation fluences. 
The observation that $V_{gl}$ decreases at ever higher fluence irradiation levels is a consequence of the loss of effective boron doping in the gain layer. 
Fig.~\ref{fig:423}(right) shows that the IHEP-IMEv2 LGAD sensors with 0.5~a.u. carbon implantation and a long-time carbon annealing recipe (W7-II) are the most radiation-robust. 
These sensors have an acceptor removal constant of 1.27$\times$10$^{-16}$ cm$^{2}$, around 3--5 times lower than the ones typically measured for boron only devices \citep[e.g.,]{wada2019HPK,Padilla_2020_DeepGainLayer,Mengzhaov1,ARCIDIACONO2020FBK,HPK_Acceptor_Removal_UFSD}.
\begin{equation}\label{eq:c}
V_{gl}\left(\phi_{eq}\right)=V_{gl}(0) e^{-c\cdot \phi_{eq}}
\end{equation}

\begin{figure}[!b]
	\centering
	\includegraphics[scale=0.36]{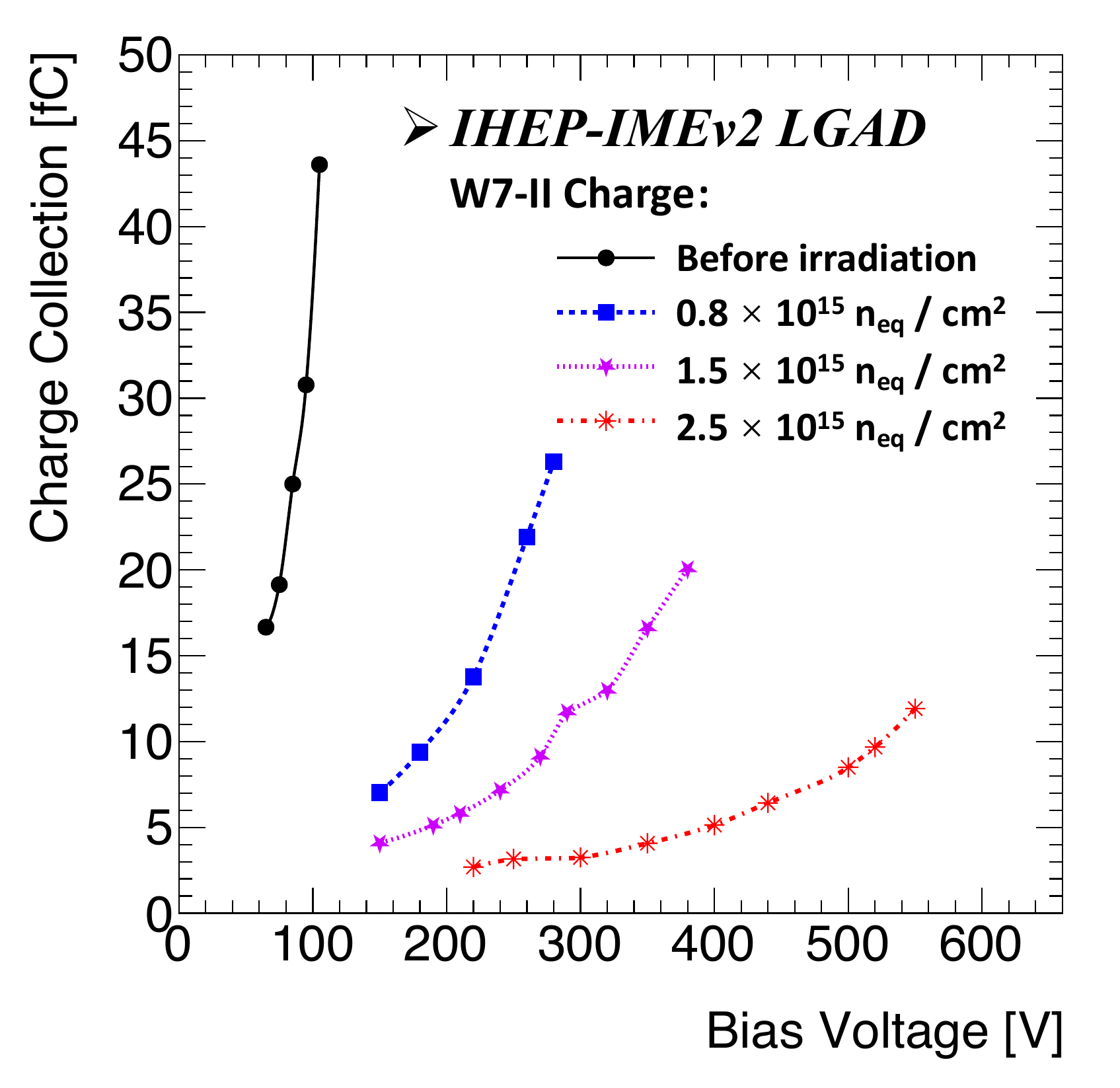}
	\includegraphics[scale=0.36]{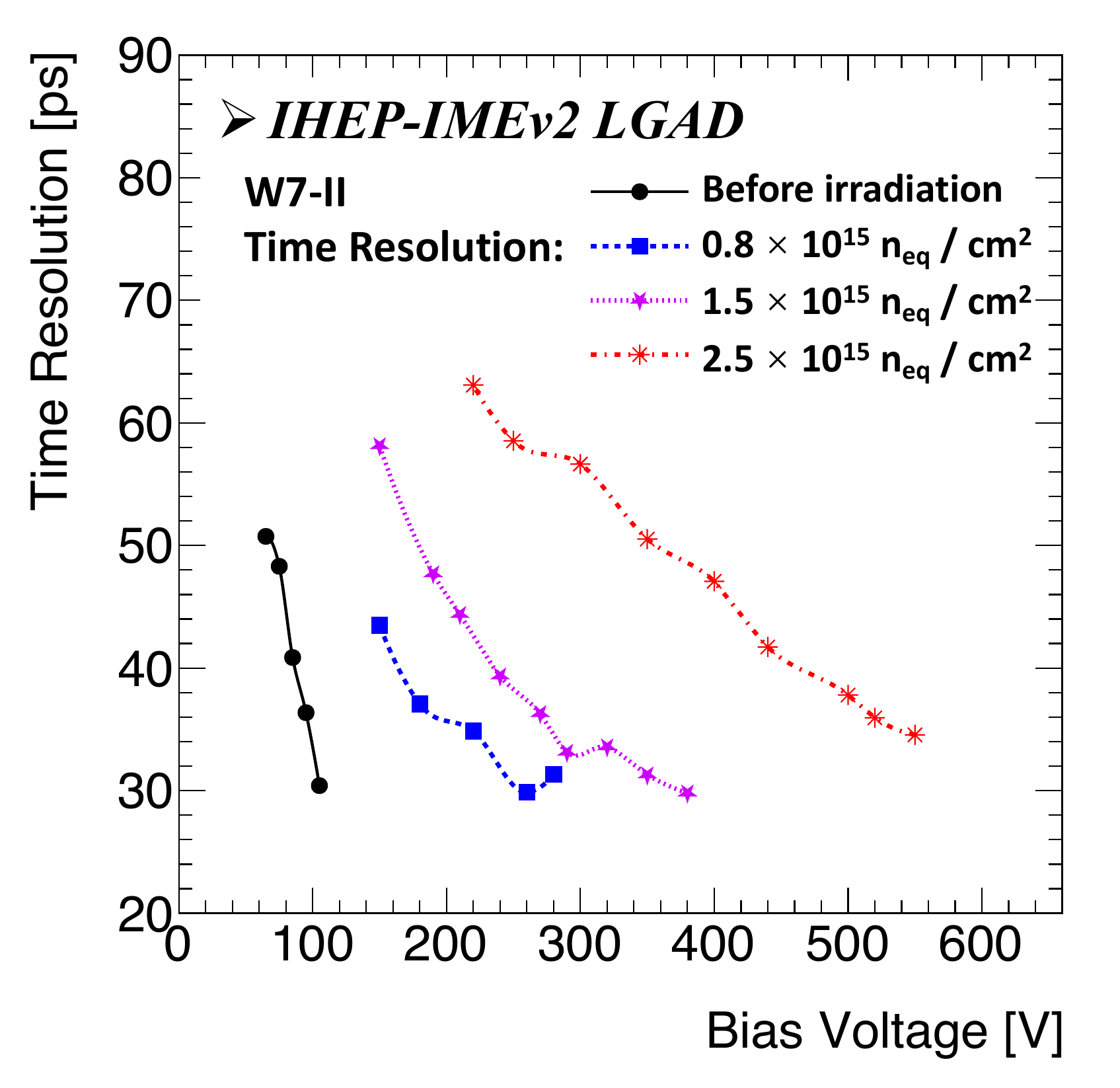}
	\caption{\label{fig:421} Collected charges (left) and time resolutions (right) of IHEP-IMEv2 W7-II LGAD sensors before radiation and after experiencing a 0.8$\times$10$^{15}$, 1.5$\times$10$^{15}$, and 2.5$\times$10$^{15}$ n$_{eq}$/cm$^{2}$ irradiation fluence. The bias voltage needed for maintaining the same charge and time resolution becomes higher. The bias voltage for 4~fC charge collection increases from $< 80$~V before irradiation to 350~V after 2.5$\times$10$^{15}$ n$_{eq}$/cm$^{2}$ irradiation.}
\end{figure}

The acceptor removal and loss of gain is the most serious consequence of radiation damage at HL-LHC. 
These can be compensated to some extent by increasing the bias voltage, but recent studies \citep{SPA_Mortality_talk_38th_RD50,Fermilab_Mortality_talk_38th_RD50} showed that the maximum bias voltage that the sensors can withstand reliably in the beam is limited to about 550~V for 50~$\mu$m thick devices, due to single event burnout (SEB). 
The charge collection efficiency of the W7-II samples and their timing performance are shown in Fig.~\ref{fig:421}. 
It is clear that the required bias for successful operation (\textgreater4 fC, \textless50 ps) increases with increasing irradiation fluence. 
However, for the W7-II sensors after receiving a 2.5$\times$10$^{15}$ n$_{eq}$/cm$^{2}$ irradiation, the required performance can be reached at a bias voltage of \textless400~V, lower than any other LGAD device studied thus far.
Apart from the obvious benefit of avoiding operation close to SEB voltage, reduced operation bias leads to smaller power dissipation.

\begin{figure}[!b]
	\centering
	\includegraphics[width=0.99\textwidth]{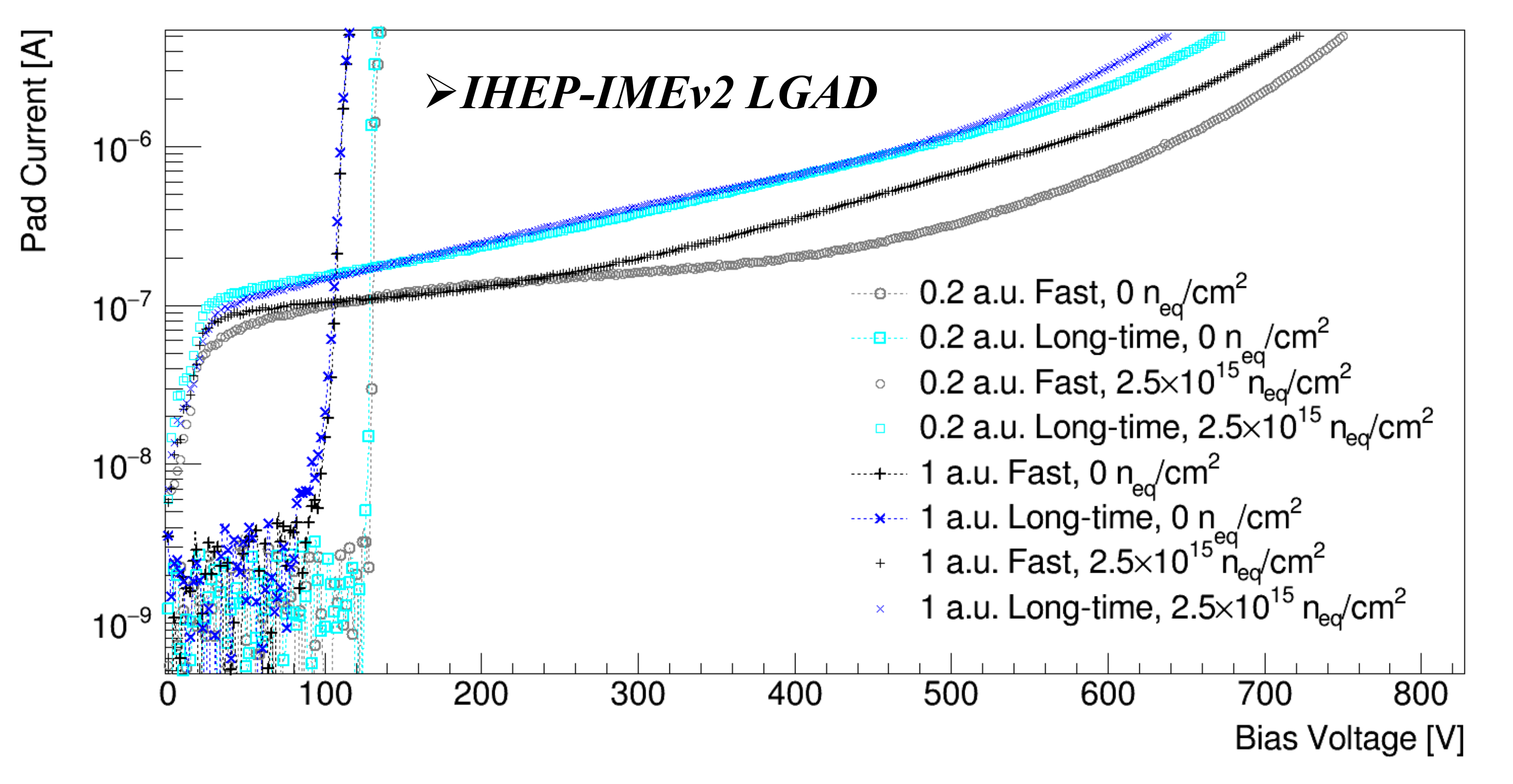}
	\caption{\label{fig:430} I--V measurements of IHEP-IMEv2 W4-I (0.2~a.u.; fast annealing), W7-I (0.2~a.u.; long-time annealing), W4-II (1~a.u.; fast annealing), and W7-III (1~a.u.; long-time annealing) LGAD sensors after receiving a 2.5$\times$10$^{15}$ n$_{eq}$/cm$^{2}$ irradiation fluence. The leakage current increases from 10$^{-9}$~A to 10$^{-7}$~A.}
\end{figure}

\subsection{Carbon thermal process}
\begin{figure}[!t]
	\centering
	\includegraphics[scale=0.36]{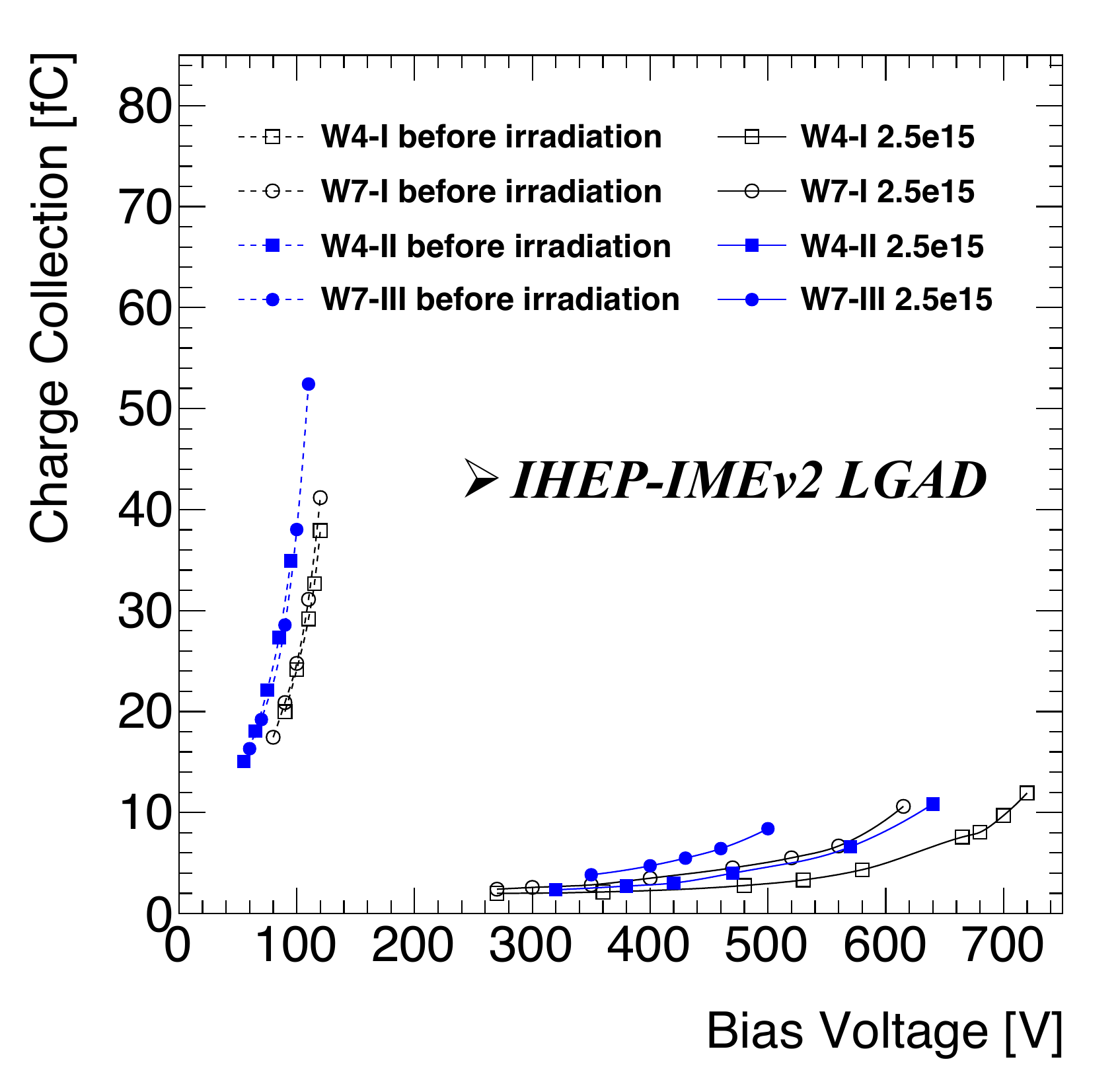}
	\includegraphics[scale=0.36]{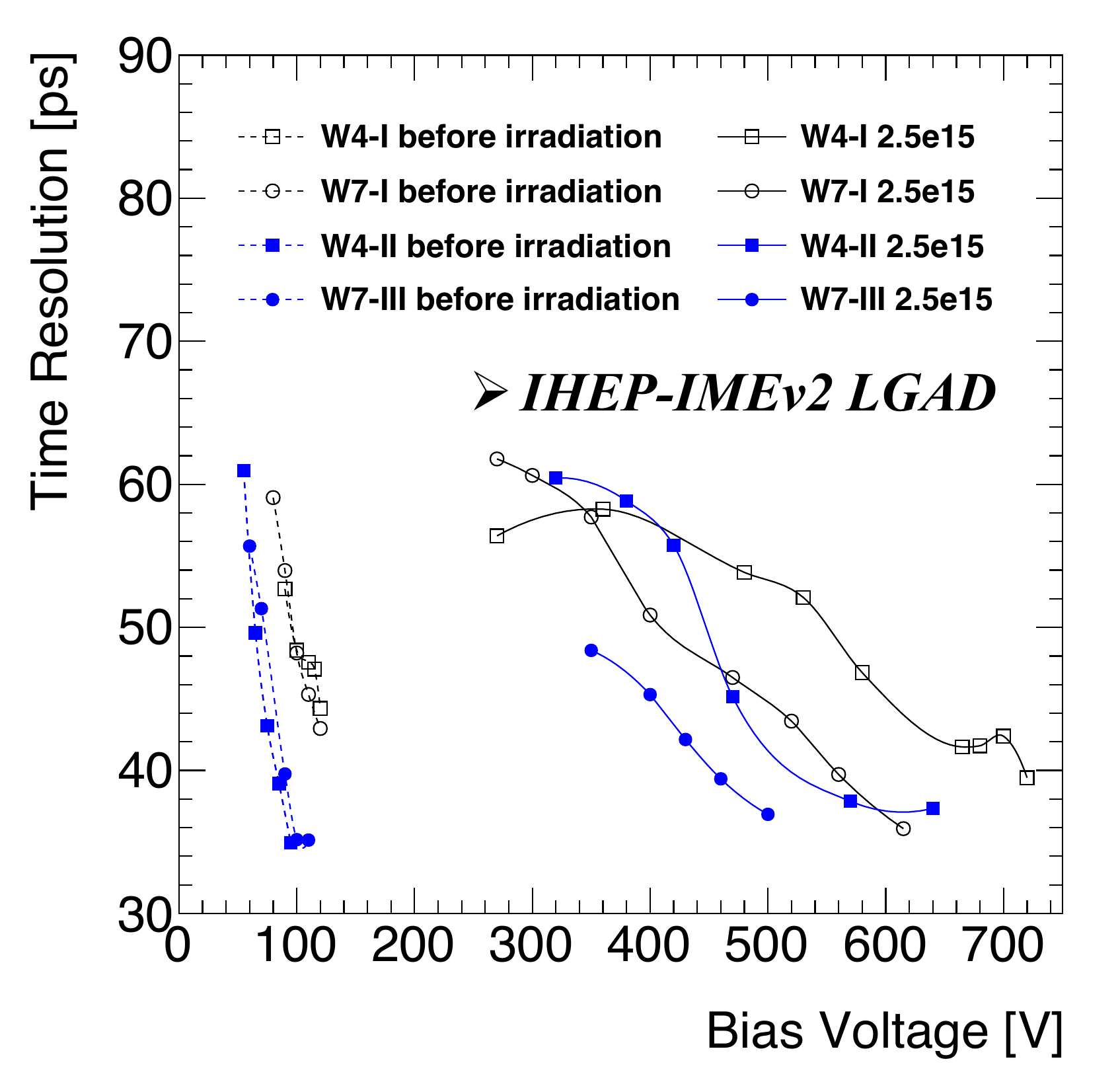}
	\caption{\label{fig:431} Most probable collected charge (left) and time resolution (right) of IHEP-IMEv2 sensors before irradiation and after receiving a 2.5$\times$10$^{15}$ n$_{eq}$/cm$^{2}$ irradiation fluence, for different thermal processes. Sensors with the same carbon dose (0.2~a.u. in black; 1~a.u. in blue) have similar behavior before irradiation, regardless of the thermal process. The long-time carbon annealing LGAD sensors (W7-I and W7-III) show better charge and time resolution after receiving a 2.5$\times$10$^{15}$ n$_{eq}$/cm$^{2}$ irradiation fluence.}
\end{figure}

In IHEP-IMEv2 production, the W4-I and W7-I (0.2~a.u.), W4-II, and W7-III (1~a.u.) wafer quadrants are good control groups to investigate the influence of the thermal process on the diffusion and activation of carbon.
\hbox{Fig.~}\ref{fig:430} and Fig.~\ref{fig:431} show the I--V, collected charge, and time resolution measurements of those two control groups.
After receiving a 2.5$\times$10$^{15}$ n$_{eq}$/cm$^{2}$ irradiation fluence, the long-time carbon annealing process (wafer 7) yields sensors with the same charge collection and time resolution as the fast annealing process (wafer 4) but at a lower bias voltage. 
This is the result of a larger thermal load delivered to the carbon, resulting in more of it diffusing into the gain layer. 
The same conclusion can also be verified by Fig.~\ref{fig:423}(right) which shows wafer 7 has a smaller acceptor removal constant for the same carbon dose as wafer 4.

\subsection{Carbon implantation dose}
\begin{figure}[!t]
	\centering
	\includegraphics[scale=0.36]{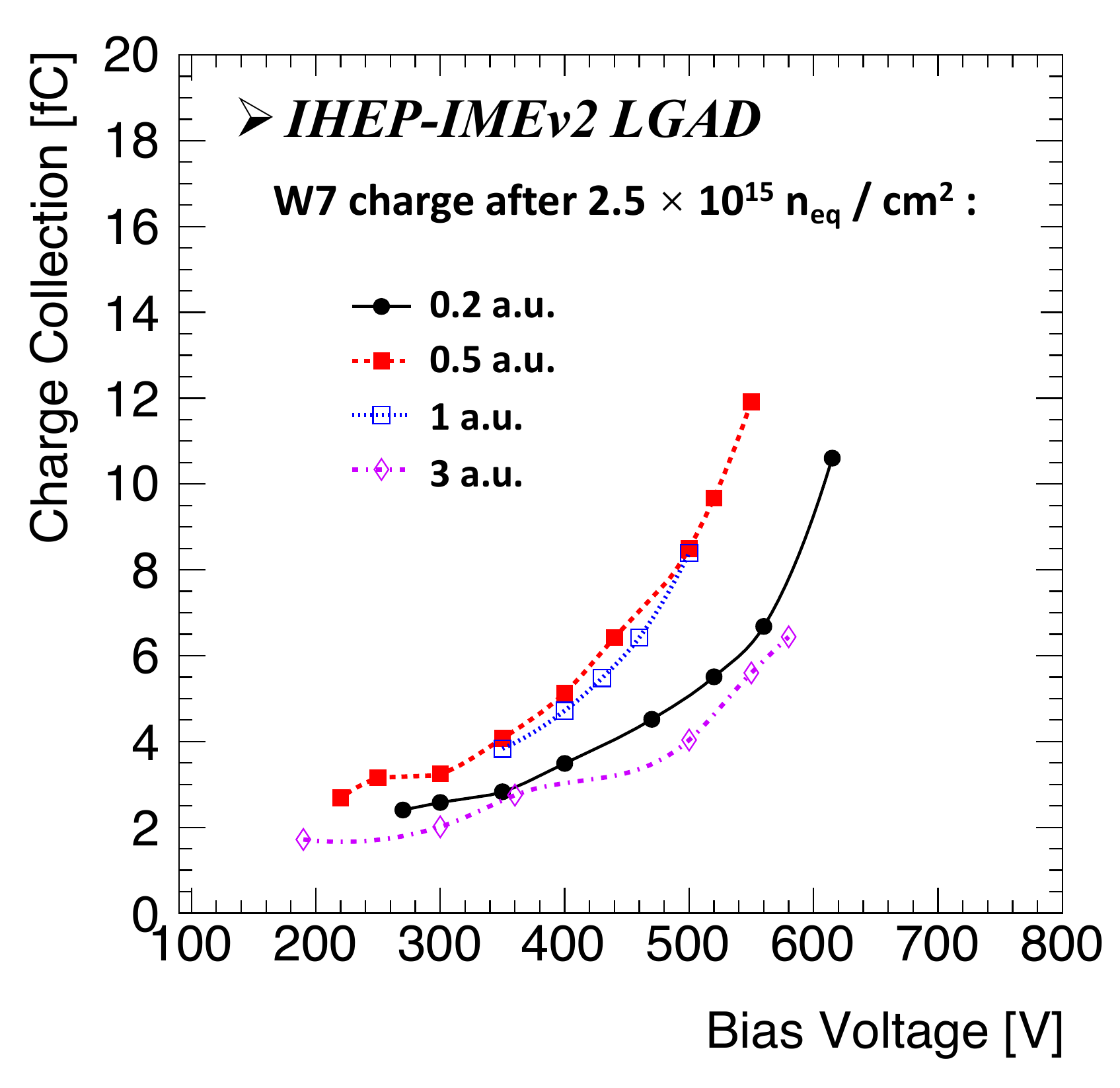}
	\includegraphics[scale=0.36]{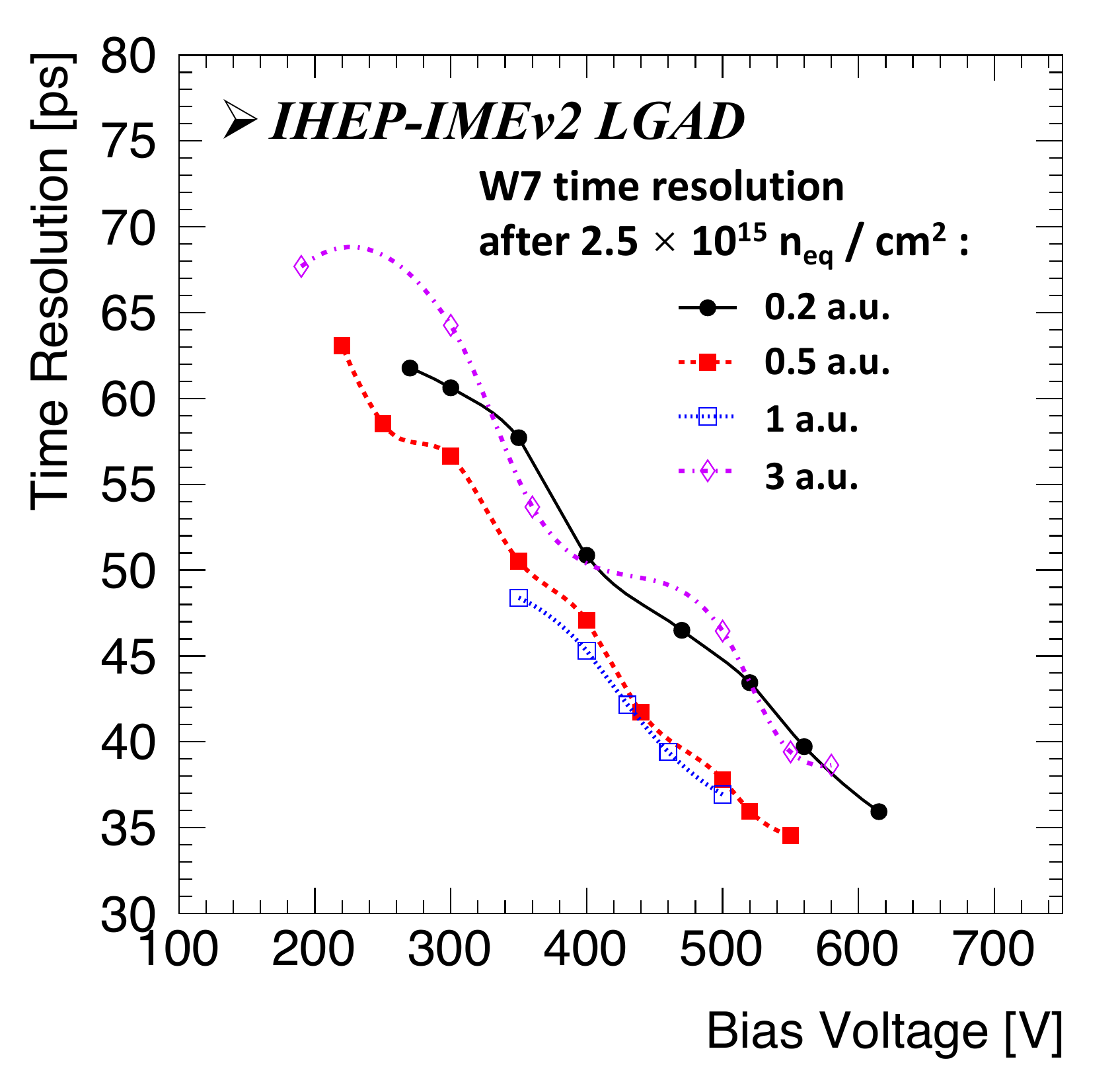}
    \caption{\label{fig:441} Most probable collected charge (left) and time resolution (right) of IHEP-IMEv2 sensors with long-during carbon annealing process (W7) after receiving an irradiation fluence of 2.5$\times$10$^{15}$ n$_{eq}$/cm$^{2}$. The red curve (0.5~a.u.) shows the best charge and time resolution performance after receiving the 2.5$\times$10$^{15}$ n$_{eq}$/cm$^{2}$ radiation dose.}
\end{figure}

It is obvious that long-time carbon annealing, as used for wafer 7, is a better process for 50~keV carbon implantation.
Fig.~\ref{fig:441} shows the collected charge and time resolution of sensors with different carbon doses (0.2, 0.5, 1, and 3~a.u.), after experiencing an irradiation fluence of 2.5$\times$10$^{15}$ n$_{eq}$/cm$^{2}$. 
The sensors with a 0.5~a.u. carbon dose (W7-II) have the best performance after irradiation, which achieve the same charge collection and time resolution as the other carbon dose sensors but at a lower bias voltage. 
These sensors also have excellent performance before irradiation, as shown in Fig.~\ref{fig:421}.

\section{Summary}

The IHEP-IMEv2 carbon enriched LGAD sensors show excellent robustness to radiation. 
In particular, measurements of leakage current level, $V_{gl}$, acceptor removal constant, collected charge, and time resolution before and after irradiation demonstrate that LGAD sensors (W7-II) produced with 0.5 a.u. carbon dose and long-time carbon annealing have excellent performance and superior irradiation resilience. 
These sensors have the smallest acceptor removal constant (1.27$\times$10$^{-16}$ cm$^{2}$) \ref{tab:0}, the lowest bias voltage (400 V) for ~50 ps time resolution, and the lowest bias voltage (350 V) for 4 fC charge collection after irradiation fluence of 2.5$\times$10$^{15}$ n$_{eq}$/cm$^{2}$ when compared to past HPK, FBK, CNM, NDL, and USTC LGAD sensors \cite{Joao_HGTD_news}. 
These sensors already satisfy the requirements for operation in the radiation harsh environment of the HL-LHC.

\section*{Acknowledgment}
This work was supported by the National Natural Science Foundation of China, under Grant No.12175252 and No.12188102; the Scientific Instrument Developing Project of the Chinese Academy of Sciences (Grant No.ZDKYYQ20200007); the State Key Laboratory of Particle Detection and Electronics, China, under project SKLPDE-ZZ-202001 and project SKLPDE-ZZ-201911; and project ARRS J1-1699 of the Slovenian Research Agency.


\end{document}